\documentclass[12pt]{iopart}
\usepackage{graphicx}
\usepackage{epsfig}
\usepackage{harvard}
\usepackage{iopams}
\usepackage{xspace}
\usepackage{lineno}
\usepackage{color}

\newcommand{\ccmskev}{\ensuremath{\text{counts}\,\text{cm}^{-2}\,\text{s}^{-1}\,\text{keV}^{-1}}\xspace}
\newcommand{\text}{\rm }

\newcommand{\CERN}{European Organization for Nuclear Research (CERN), CH-1211 Gen\`eve 23, Switzerland}
\newcommand{\Saclay}{DAPNIA, Centre d'\'Etudes Nucl\'eaires de Saclay (CEA-Saclay), Gif-sur-Yvette, France.}
\newcommand{\Zaragoza}{Laboratorio de F\'{\i}sica Nuclear y Altas Energ\'{\i}as, Universidad de Zaragoza, Zaragoza, Spain }
\newcommand{\Vancouver}{Department of Physics and Astronomy, University of British Columbia, Vancouver, Canada }
\newcommand{\Zagreb}{Rudjer Bo\v{s}kovi\'{c} Institute, Zagreb, Croatia}
\newcommand{\Pisa}{Scuola Normale Superiore, Pisa, Italy.}
\newcommand{\Lyon}{Inst. de Physique Nucl\'eaire, Lyon, France.}
\newcommand{\queens}{Department of Physics, Queen's University, Kingston, Ontario K7L 3N6, Canada}
\newcommand{\patras}{University of Patras, Patras, Greece}
\begin{document}

\title{The CAST Time Projection Chamber}

\author{
D~Autiero$^{1,}$\footnote{Present addr.: \Lyon},  
B~Beltr\'an$^{2}$, 
J~M~Carmona$^3$, 
S~Cebri\'an$^3$, 
E~Chesi$^1$,
M~Davenport$^{1}$,
M~Delattre$^1$,
L~Di~Lella$^{1,}$\footnote{Present addr.: \Pisa}, 
F~Formenti$^1$,
I~G~Irastorza$^{1,}$\footnote{Present addr.: \Saclay},
H~G\'omez$^3$,
M~Hasinoff$^{4}$,
B~Laki\'c$^{5}$,
G~Luz\'on$^{3}$,
J~Morales$^{3}$,
L~Musa$^1$,
A~Ortiz$^{3}$,
A~Placci$^{1}$,
A~Rodriguez$^3$, 
J~Ruz$^{3}$, 
J~A~Villar$^{3}$ and
K~Zioutas$^{1,6}$
}

\address{$^{1}$ \CERN}
\address{$^{2}$ \queens}              
\address{$^{3}$ \Zaragoza}
\address{$^{4}$ \Vancouver}
\address{$^{5}$ \Zagreb}              
\address{$^{6}$ \patras}              
\ead{jaime.ruz@cern.ch}

\begin{abstract}
  One of the three X-ray detectors of the CAST experiment searching for
  solar axions is a Time Projection Chamber (TPC) with a multi-wire
  proportional counter (MWPC) as a readout structure. Its design has been
  optimized to provide high sensitivity to the detection of the low
  intensity X-ray signal expected in the CAST experiment. A low hardware
  threshold of 0.8~keV is safely set during normal data taking periods, and
  the overall efficiency for the detection of photons coming from
  conversion of solar axions is 62\%.  Shielding has been installed around
  the detector, lowering the background level to
  4.10~$\times10^{-5}\ccmskev$ between 1 and $10\,\text{keV}$.  During
  phase~I of the CAST experiment the TPC has provided robust and stable
  operation, thus contributing with a competitive result to the overall
  CAST limit on axion-photon coupling and mass.
\end{abstract} 

\submitto{\NJP}


\section{Introduction}
The CERN Solar Axion Telescope (CAST) is the most sensitive implementation
of the ``helioscope'' concept to look for hypothetical axions (or
axion-like particles) coming from the Sun \cite{sikivie:83a}. CAST makes
use of a decommissioned LHC prototype magnet with a length of
$9.3\,\text{m}$, providing a $9\,\text{Tesla}$ field to trigger the
conversion of solar axions into X-rays, which can then be detected by the
three different X-ray detectors placed at both ends of the magnet
\cite{zioutas:99a}. The purpose of the present paper is to describe one of
the detector systems of CAST: the time projection chamber (TPC).
Accompanying papers are devoted to the other two X-ray detectors
\cite{kuster:06a,andriamonje:06a} as well as to an overall description of
the experiment \cite{andriamonje:07b}. The first physics results of CAST
phase~I have been published in \citeasnoun{zioutas:05a} and
\citeasnoun{andriamonje:07a}.

The TPC detector is attached to eastern end of the CAST magnet, covering
both magnet bores and being therefore exposed to the converted photons from
``sunset'' axions during evening solar tracking. The design of the detector
follows the well-known concept of a conventional TPC, i.e., a large volume
gaseous space where primary interactions take place, producing ionization
electrons which drift towards a plane of wires. Here, as in a Multi Wire
Proportional Chamber (MWPC), the avalanche process that amplifies the
signal is developed, allowing a position sensitive readout of the original
event. The specific requirements of the CAST experiment (sensitivity to a
low intensity X-ray signal peaking at $\sim 4\,\text{keV}$ and vanishing at
around $10\,\text{keV}$), necessitated some original approaches in the
construction of the CAST TPC. An optimum performance would include: low
threshold (at the keV level), a relatively high gain, position sensitivity
(to distinguish events coming from the magnet bores and to allow some
degree of background rejection by pattern recognition), good efficiency for
the energies of interest, low background at low energies, and last but not
least, robust and stable operation over the long data taking periods needed
to accumulate enough statistics.  The CAST TPC was designed to fulfill
these requirements using conventional reliable technology.

Our paper is structured as follows: in section~\ref{sec:chamb-shield-descr}
description of the detector itself and the shielding installed is given.
It is followed by a description of the data acquisition electronics in
section~\ref{sec:data-acqu-hardw} and the characterization of the detector
based on the calibration data at the PANTER X-ray test facility in Munich
in section~\ref{sec:characterization}. Finally, the basic data treatment is
described in section~\ref{sec:data-treatment}, stressing the aspects
concerning off­line background strategies and the detector performance
during data taking period of CAST phase~I will be presented in
section~\ref{sec:detector-performance}.

\section{Chamber and Shielding Concept}
\label{sec:chamb-shield-descr}
\subsection{Chamber Description}
The CAST TPC has a conversion volume of $10\times15\times30\,\text{cm}^3$.
The $10\,\text{cm}$ drift direction is parallel to the magnet beam pipes,
and the section of $15\times30\,\text{cm}^2$ is perpendicular to that
direction, covering both magnet bores (each bore has a diameter of
$42\,\text{mm}$ and their centers are separated by $18\,\text{cm}$). This
conversion volume, filled with gas Ar(95\%)/CH$_4$(5\%) at atmospheric
pressure, allows for essentially total conversion ($>99\%$) of photons up
to $6\,\text{keV}$ crossing the chamber parallel to the magnet beam
direction. The conversion efficiency decreases for higher energies,
becoming 50\% for photons of $11.5\,\text{keV}$. The gas is being
continuously renewed at a flow rate of $2\,\text{l}\,\text{h}^{-1}$, in
order to prevent any contamination from atmospheric impurities, such as
N$_2$ or O$_2$.
\begin{figure}[t]
  \begin{center}
    \includegraphics[width=1.06\textwidth]{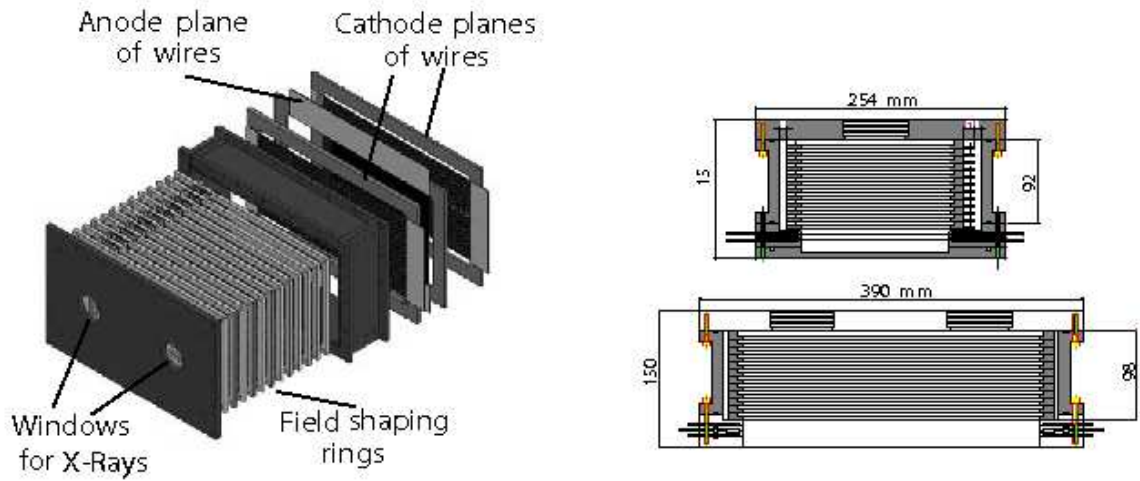}
  \end{center}
  \caption{Left: Exploded view of the TPC. The pieces holding the anode
    and cathode wire planes and the field shaping conductive frames are
    clearly seen.  The holes to hold the circular windows for the X-rays
    coming out from the magnet bores are present in the front piece.
    Right: Both side views of the TPC showing the general dimensions in
    millimeters.}
  \label{tpc-figure}
\end{figure}

The maximum drift distance is $10\,\text{cm}$ between the drift electrode
and the sense wires. The drift electrode is a continuous aluminum layer
located on the inner side of the chamber wall closer to the magnet,
extended to its whole $15\times30\,\text{cm}^2$ dimension. It is biased at
$-7\,\text{kV}$, thereby producing an electric field of
$700\,\text{V}\,\text{cm}^{-1}$. To shape the electric field at the chamber
edges, several rectangular ($15\times30\,\text{cm}^2$) conductive frames at
intermediate voltages stepping between 0 and $-7\; {\rm kV}$ are used (see
\fref{tpc-figure}). On the back side of the chamber, the wires are arranged
in 3 planes, one anode plane at $+1.8\,\text{kV}$ between two grounded
cathode planes. The anode plane contains $48$ wires of $20\mu\text{m}$
diameter (gold plated tungsten) which run parallel to the wider side of the
chamber.  Each cathode plane contains 96 wires of $100 \mu\text{m}$
diameter that run perpendicular to the anode wires.  The intense electric
field surrounding the anode (sense) wires causes the avalanche process and
hence the multiplication of the primary ionization cloud. The positive ions
produced in an avalanche event are cleared away by the neighboring cathode
wires, which receive an induced signal, providing two dimensional
information for each event. The distance between adjacent wires of the same
plane is $3\,\text{mm}$.  The distance between the anode and the cathode
plane closest to the drift region is $3\,\text{mm}$, while the distance
between the anodes and the outer cathode plane is $6\,\text{mm}$. This
asymmetric configuration enhances the induced signal on the cathode wires
closest to the drift region, which are the ones read out by the
electronics. As a result, the ratio between the anode and cathode signals
for a given energy deposition in the chamber, a parameter purely determined
by geometry, is approximatively a factor of 2 in our case.

With the exception of the electrodes themselves, plus the screws and the
Printed Circuit Board (PCB) where the wires are held, the entire chamber is
made of plexiglass.  The radioactivity level of this material has been
measured at the Canfranc Underground Laboratory facility and found to be
very low ( $< 100\,\text{mBq}\,\text{kg}^{-1}$ of $^{238}$U,
$<10\,\text{mBq}\,\text{Kg}^{-1}$ of $^{235}$U,
$<5\,\text{mBq}\,\text{kg}^{-1}$ of $^{232}$Th and
$<30\,\text{mBq}\,\text{kg}^{-1}$ of $^{40}$K) therefore suitable for our
low background application. The thickness of the plexiglass wall is about
$1.7\,\text{cm}$, except in three places (two on the back side and one on a
lateral side) where the only separation between the inner gas and the
atmosphere is a thin mylar foil to allow the calibration of the chamber
with low energy X-ray sources.  In addition, the side closet to the magnet
has two $6\,\text{cm}$ radius circular holes for the thin windows that must
be as transparent as possible to the X-rays coming through the magnet,
while being able to hold the large pressure difference ($1\,\text{atm}$)
between the chamber gas and the magnet bore.

These windows are basically very thin mylar foils ($3$ or $5\,\mu\text{m}$)
stretched, then glued to a metallic grid (strongback) on the vacuum side of
the foil. This technique allows the thin foil to withstand the large
pressure difference.  The geometrical opacity of the strongback is about
8\% while the mylar foil is practically transparent for X-rays down to the
keV energies \citeaffixed{henke:93a}{$\sim 30\%$ transparency for $1\,{\rm
    keV}$, $\sim85\%$ for $2\,\text{keV}$ and $\sim95\%$ for
  $3\,\text{keV}$, see}.  The inner side of the mylar foil is aluminized
($40\,\text{nm}$) since it serves as part of the drift electrode. To
minimize the effect of the chamber gas leaks towards the magnet, and
therefore to cope with the stringent requirements of the magnet vacuum
system, a differential pumping system has been installed.  This system
creates an intermediate volume between the TPC and the magnet which is
continuously being pumped with a clean pump. This volume is kept at a
relatively poor vacuum ($\sim10^{-5}\,\text{mbar}$, compared with
$\sim10^{-7}\,\text{mbar}$ in the magnet). A second thin polypropylene
window separates this intermediate volume from the magnet vacuum. Due to
the small pressure difference, the effective leak through this window is
extremely small ($1.46\times10^{-7}\,\text{mbar}\,\text{l}\,\text{s}^{-1}$
of Argon). This strategy allows us to be reasonably tolerant to small leaks
on the TPC windows, improving the robustness of the whole system.

\subsection{Background Sources and Shielding Concept}
\label{sub:back and shileding}
The CAST experiment is located in one of the surface buildings of the PA8
experimental area at CERN. Due to the fact that the TPC is attached to the
end of the magnet which is far from the pivot point, it moves not only over
relatively large distances during each daily solar tracking but also shifts
position within the hall progressively following the seasons. If the
spatial distribution of possible background sources is not uniform in the
hall, this movement may induce background fluctuations in the detector.

To assess possible sources of background for the TPC, one has to take into
account the high event discrimination capabilities of the TPC. Most of the
events contributing to the raw trigger rate are muons and other cosmic-ray
related, multi-cluster (or track) events, and they are easily rejected by
the software cuts described in section~\ref{sec:data-treatment}. In the
following, we denote as background only those events which survive the
cuts, i.e., which leave a single point-like energy deposition in the
detector.

Events which contribute to this remaining background are photons induced by
cosmic rays in the surrounding materials or produced by the radioactive
chains and potassium in laboratory floor, building and experimental
materials~\cite{dumont}.  The TPC detector is practically blind to most of
this high energy radiation, but low energy Compton interactions in the
detector are still possible. Also, their interaction in the surrounding
materials can produce a diffuse gamma background of lower energies, with a
high probability of interacting in the gas. These components depend on the
concentration of the original emitters in the laboratory floor and walls
and, in general, on the geometry of the detector environment, therefore
causing a non-uniform background. Radon is also present everywhere and its
concentration has been measured to vary widely between $5$ and
$50\,\text{Bq}\,\text{m}^{-3}$ depending on causes such as walls or floor
proximity, ventilation, and the atmospheric pressure, temperature and
humidity.  The neutron component of the background is below the level of
the typical gamma background by three or four orders of magnitude, but
neutron signal in the detector could mimic those from X-rays. The dominant
sources of this background are cosmic rays, but neutrons can be also
induced by muons in surrounding materials, by ($\alpha$,~n)~reactions on
light elements and by spontaneous fission.
\begin{figure}[t]
  \begin{center}
    \includegraphics[width=0.5\textwidth]{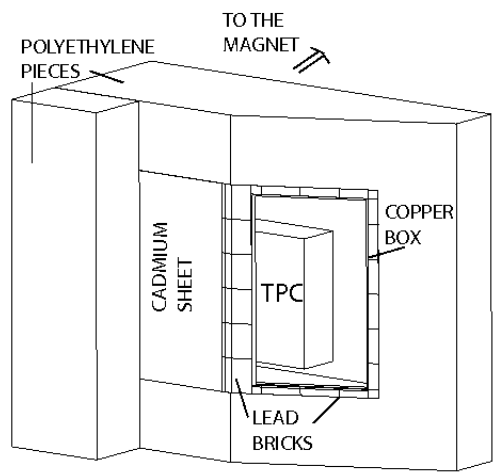}
  \end{center}
  \caption{Shielding scheme showing all the layers that surround the
    TPC, from the inside to the outside: a copper box, a lead , cadmium,
    and polyethylene shield.}
  \label{shielding}
\end{figure}

In view of these general considerations, a shielding around the TPC was
designed not only to reduce the general level of background seen by the
detectors, but also to homogenize its position dependence. The final design
is the outcome of several simulations and experimental tests, some of which
have been performed in the actual TPC to assess the effect of the shielding
on the detector background data \cite{ruz:05a}. Also a compromise was to be
found between the shielding effect and the weight and size restrictions
imposed by the limitations of the lifting screws and support structure.
From the inside to outside, the CAST TPC shielding (see \fref{shielding})
is composed of:
\begin{itemize}
\item {Copper box, $5\,\text{mm}$ thick:} it reduces the electronic noise,
  as a Faraday cage, and stops low energy X-rays produced in the outer part
  of the shielding by environmental gamma radiation. It is also used for
  mechanical support purposes.
\item {Lead wall, $2.5\,\text{cm}$ thick:} To reduce the low and medium
  energy environmental gamma radiation.
\item {Cadmium layer, $1\,\text{mm}$ thick:} to absorb the thermal neutrons
  slowed down by the outer polyethylene wall.
\item {Polyethylene wall, $22.5\,\text{cm}$ thick:} to slow the medium
  energy environmental neutrons down to thermal energies.
\item {PVC bag:} to cover the whole shielding assembly. This tightly closes
  the entire set-up allowing to flush the inner part with pure N$_2$ gas
  (coming from liquid nitrogen evaporation) at an average rate of
  $250\,\text{l}\,\text{h}^{-1}$, in order to remove radon from this space.
\item {Scintillating veto:} placed at the top of the shielding, a
  scintillating plastic of $80 \times 40 \times 5\,{\rm cm}^3$ is placed to
  reject muon-induced events by anti-coincidence with the detector.
\end{itemize}

\section{Data Acquisition Hardware and Software}
\label{sec:data-acqu-hardw}
The first stage of wire signal amplification and shaping is performed by 36
(12 for the anodes and 24 for the cathodes) ALCATEL SMB302 4-channel
preamplifier chips, located on the same printed circuit board on which the
wires are supported. The output from these preamplifiers is sampled by
three 48-channel 10-bit VME flash-ADCs operating at a sampling rate of 10
MHz. These modules are based on the ALTRO (ALICE TPC Read-Out) chip,
developed at CERN for the ALICE experiment \cite{bosch:03a}. The same
modules have been used in the HARP and CERES (NA45) experiments at CERN.

The trigger is built from the OR of all the anode signals. The hardware
trigger threshold during normal CAST data-taking operation is safely set to
avoid being triggered by electronic noise, and corresponds to energy
depositions of about $800\,\text{eV}$ in the gas conversion region.  The
time window for the sampling is about $7\,\mu\text{s}$, which is long
enough to encompass the maximum drift time of the chamber. The flash-ADCs
are configured, controlled and read through a VME bus controlled by a dual
processor PC running under Linux, which uses a SBS Bit-3 1003 adapter on a
fiber-optic link.

The time spent by the system after the trigger arrival in
hardware-processing the data, transferring them to the PC and writing them
to the disk is, on average, about $1.5\,\text{ms}$, during which it is
``blind'' to new triggers. This means that for the typical chamber trigger
rate of $10$ to $25\,\text{Hz}$ --depending of data taking conditions-- the
average dead time can vary from $1.75 \%\;{\rm to}\; 2.5 \%$.  The dead
time is continuously monitored by measuring the time the system spends in
the ``BUSY'' state, defined by a hardware register which is set by the
trigger and reset to zero by the PC once the event processing is finished.
The BUSY logic is handled by a CORBO VME module. The dead time is therefore
calculated on-line by using a scaler to count clock pulses both with and
without a veto from the BUSY signal provided by the CORBO module.

The acquisition software is a low level C-code which configures,
initializes and controls the electronics modules through the VME bus. Once
a trigger is detected, its task is basically to dump the contents of the
flash-ADC memories onto the disk, without any further data treatment, so as
not to add any dead time. A different, high level C-code for passive
monitoring of the detector performance is running continuously in the
second processor, without interfering with the acquisition (and therefore
without adding noticeable dead time). This software, based on the ROOT
toolkit developed for data analysis at CERN \cite{brun:97a}, monitors
online multiple experimental parameters, and therefore allows fast
diagnosis of problems and helps in assessing the quality of the data as
they are being acquired.

The data acquisition protocol follows a fully automatic procedure, and data
belonging to Sun-tracking (``axion-sensitive'') measurements or to
background measurements are identified and separated during the off-line
analysis (by using the slow control data). Therefore, the acquisition
starts and stops are set only by practical reasons, mainly to allow a
calibration of the detector. The gain of the detector is measured using a
$^{55}$Fe source and the pedestal levels and the variation of the
flash-ADCs channels are measured using artificial trigger signals in the
absence of real events. As a primary acquisition protocol, the acquisition
is automatically stopped every 6 hours to take a pedestal run and two
calibration runs (one through each back window). Then the normal
acquisition is resumed. The movement of the calibration source from the
``shielded'' parking position to the corresponding window is performed by a
stepping motor, which is fully controllable via ECL or TTL signals. These
signals are provided by an input/output register VME module controlled by
the acquisition software, so the whole acquisition sequence is fully
automated.

\section{Characterization}
\label{sec:characterization}
To precisely determine the detection properties (efficiency, gain
linearity, etc.) of the TPC, it was transported and mounted at the PANTER
facility of the Max-Planck-Institut f\"ur extraterrestrische Physik (MPE)
in Munich in 2002 \cite{freyberg:06a}. This facility, designed for the
calibration and characterization of X-ray telescopes, provides a parallel
X-ray beam with a very accurately calibrated energy and intensity.  The
data obtained in this facility have been used to determine the efficiency
of the TPC, as well as other experimental parameters such as the energy
resolution and the linearity of the detector response over the whole energy
range of interest.  These calibration data have also been used to determine
the efficiency loss in the off-line analysis of the data, in particular the
energy dependence of the off-line cuts which are applied to reduce the
background.

\begin{figure}
  \begin{center}
    \begin{minipage}{0.49\textwidth}
      \includegraphics[width=1\textwidth]{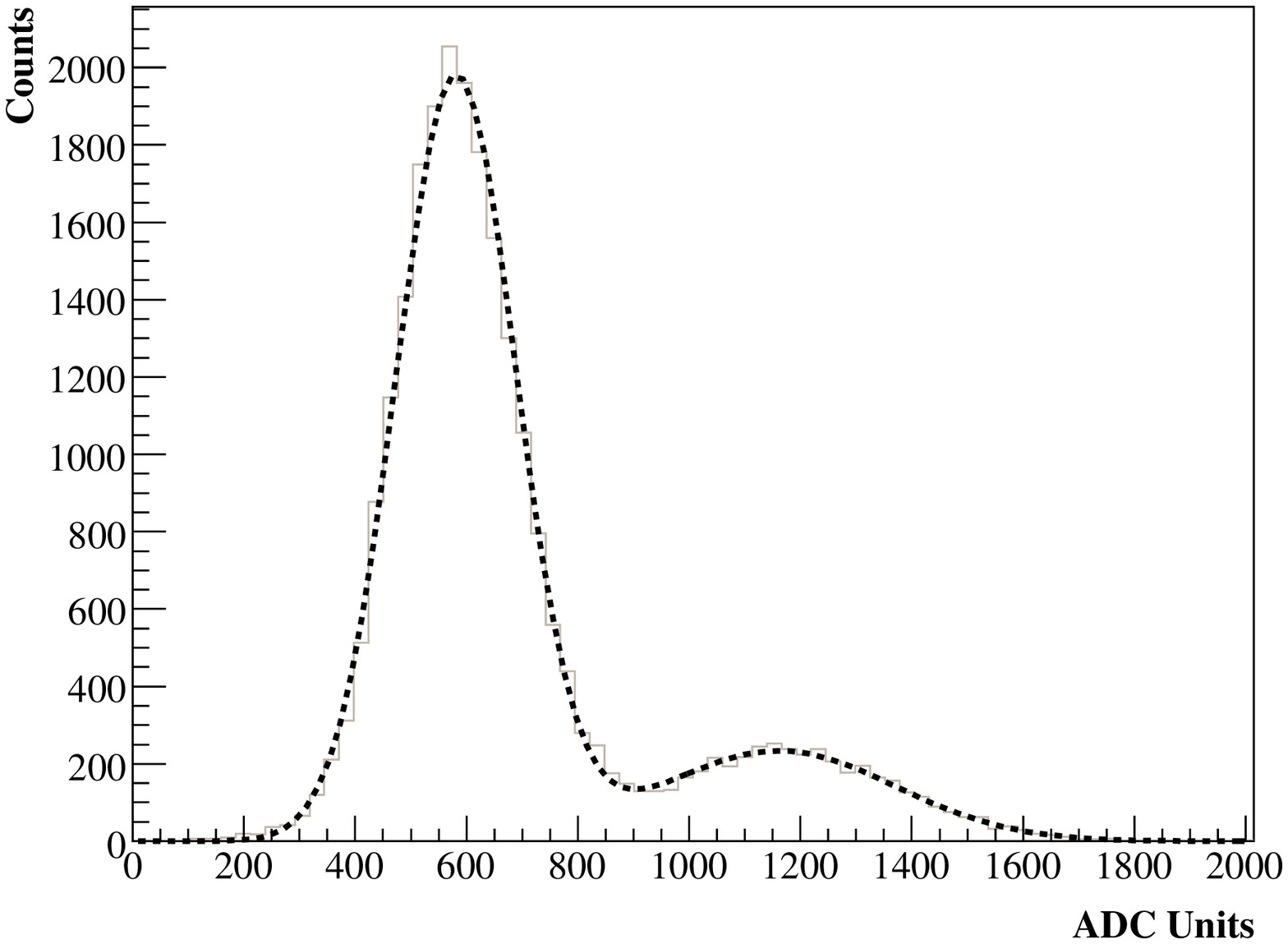}
    \end{minipage}
    \begin{minipage}{0.49\textwidth}
      \includegraphics[width=1\textwidth]{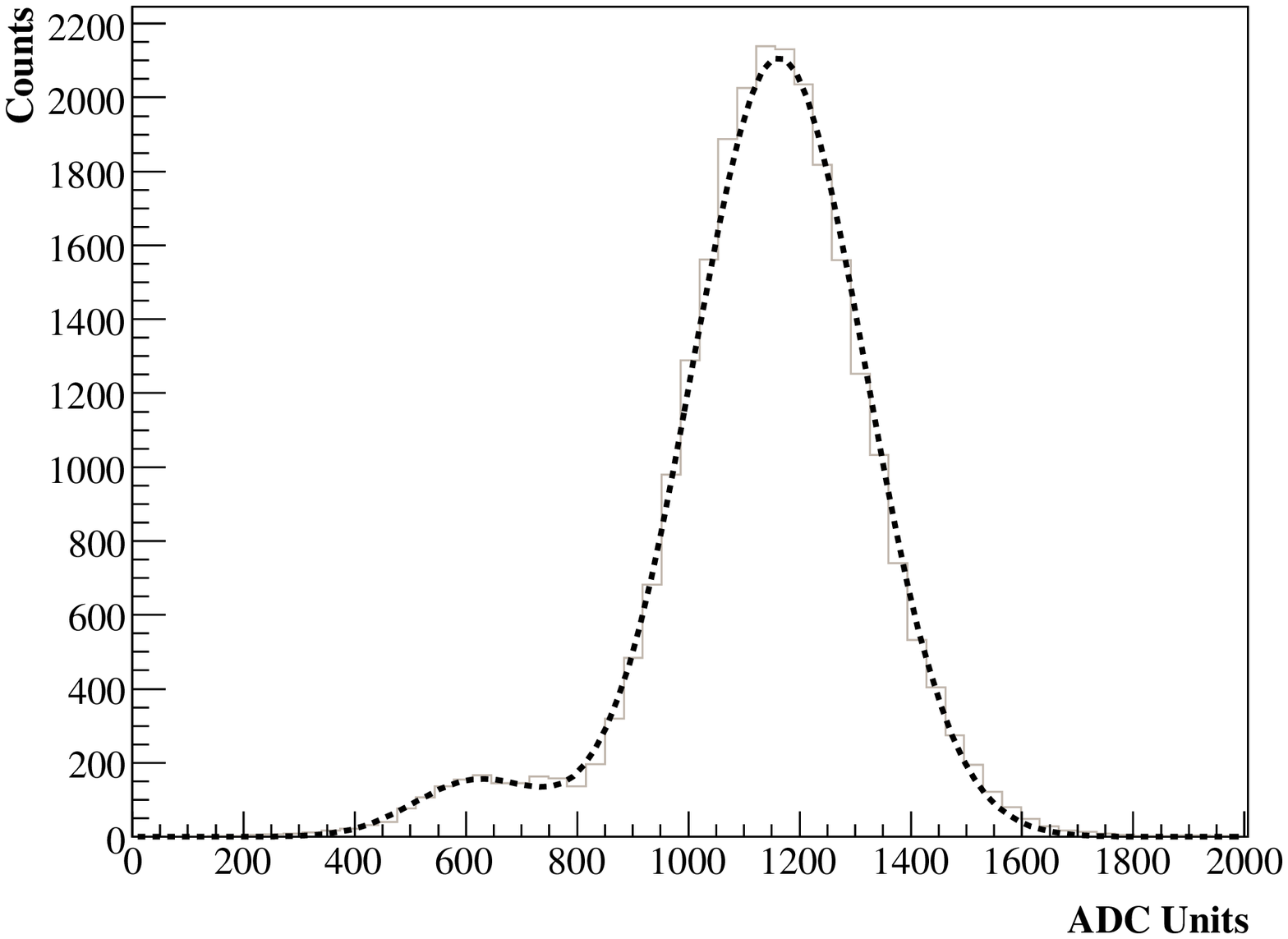}
    \end{minipage}
  \end{center}
  \caption{TPC X-ray calibration spectra for an energy of $3\,{\rm keV}$
    and $6.4\,{\rm keV}$ (Fe-K line, right) measured at the PANTER test
    facility. In the Fe-K spectrum a second $3.4\,\text{keV}$ escape peak
    is present. In the left spectrum, the bremsstrahlung continuum appears
    at higher energies in addition to the main peak. The presence of a
    small amount of bremsstrahlungs background can be deduced from the
    counts to the left of the peak. In both figures the line shows the
    combined fit of the peaks plus background.}
  \label{Panter_fits}
\end{figure}

The efficiency curve of the detector has been determined by comparing the
counts detected in each corresponding run with the expected rate deduced
from the calibrated PANTER detector. The energies provided by PANTER were:
0.3, 0.9, 1.5, 2.3, 3.0, 4.5, 6.4, and $8.0\,\text{keV}$. The beam was
``contaminated'' by a bremsstrahlung continuum for the low energy cases
(from $0.34$ up to $3\,\text{keV}$) and the chamber spectra show the
presence of a small amount of background. In the case of high energies
(from $4.5$ up to $8\,\text{keV}$), a second peak due to escape from the
argon gas also appears; hence the precise counts corresponding to the main
peak have been extracted by means of Gaussian line-shape fits to the
measured spectra.  Examples of such fits are given in \fref{Panter_fits}.
\begin{figure}
  \begin{center}
    \begin{center}
      \includegraphics[width=0.7\textwidth]{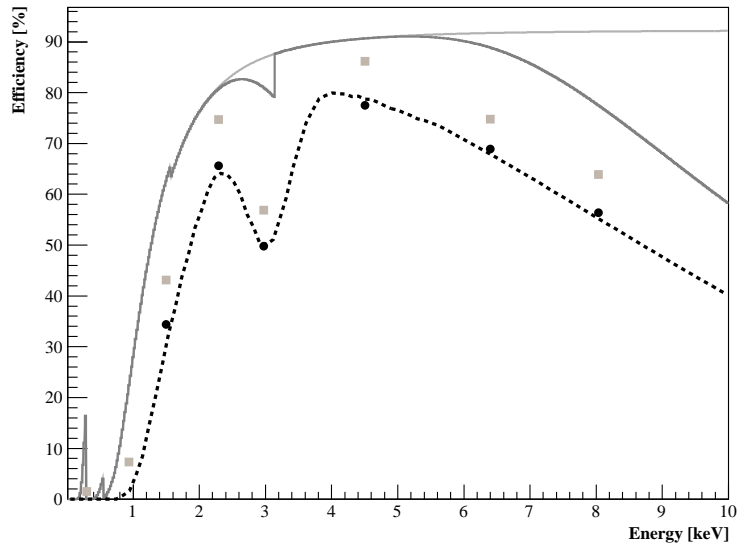}
    \end{center}
    \caption{Experimental measurements of the TPC effciency, before (grey
      squares) and after (black points) the off-line analysis cuts are
      applied to the data. The upper light grey line represents the
      theoretical computation of the window transmission, while the dark
      grey line includes also the opacity of the gas in the chamber. The
      black dashed line is the analytical function used to interpolate the
      experimental points in the final analysis.}
    \label{efficiency}
  \end{center}
\end{figure}
The results of these analyses are illustrated in \fref{efficiency}.  The
points represent the experimental values and the lines are the theoretical
computations. The dark grey line represents the expected fraction of
photons passing through the windows and being absorbed somewhere in the
sensitive gas volume of the chamber. It is therefore a theoretical (and
optimistic) prediction for the efficiency of the detector, taking into
account the two main physical effects, i.e., the window transmission and
the gas opacity to the incident X-rays. The window transmission
contribution is singled out by the upper light grey line \cite{henke:93a},
so one can easily see the contribution of both effects separately.

The experimental measurements (grey squares) for each tested PANTER energy
closely follow the values expected by the window transmission computation
for energies below $3\,\text{keV}$, lying below the grey line for energies
above $3\,\text{keV}$ due to the presence in the chamber of non point-like
events (partial or split energy depositions), which are rejected in the
off-line analysis. This loss of efficiency, in agreement with Monte Carlo
simulations, is acceptable because of the high background reduction
obtained by this approach.

The final off-line analysis (black points) produces an additional loss of
efficiency of about $5$ to $10$~\% depending on the energy. The black line
is an analytical function used to interpolate the measured efficiencies. By
multiplying this function with the expected solar axion spectrum, we obtain
an overall detection efficiency of 62\% for photons coming from conversion
of solar axions.
\begin{figure}
  \begin{center}
    \begin{minipage}{0.49\textwidth}
      \includegraphics[width=1.1\textwidth]{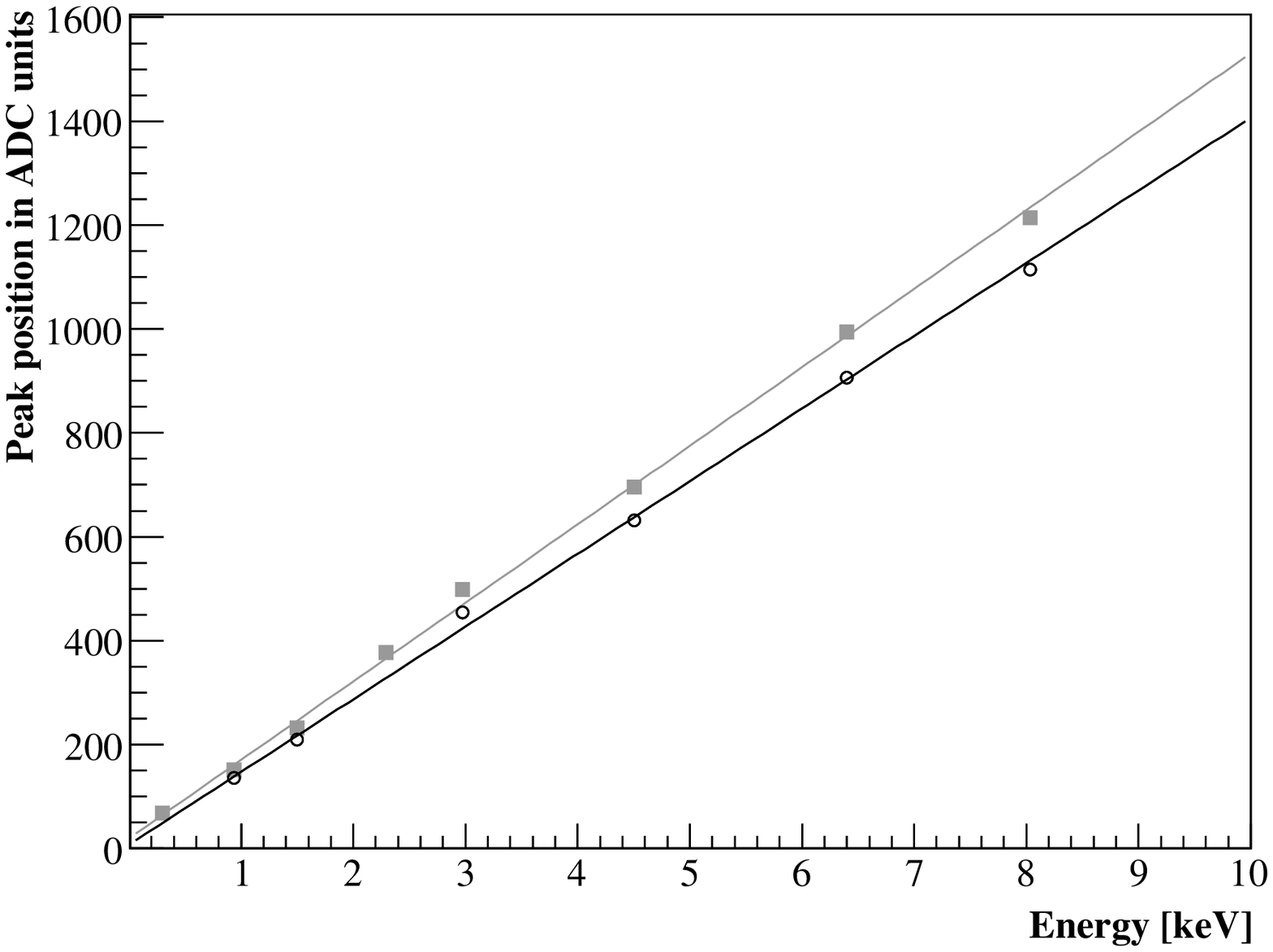} 
    \end{minipage}
    \hfill
    \begin{minipage}{0.49\textwidth}
      \includegraphics[width=1.1\textwidth]{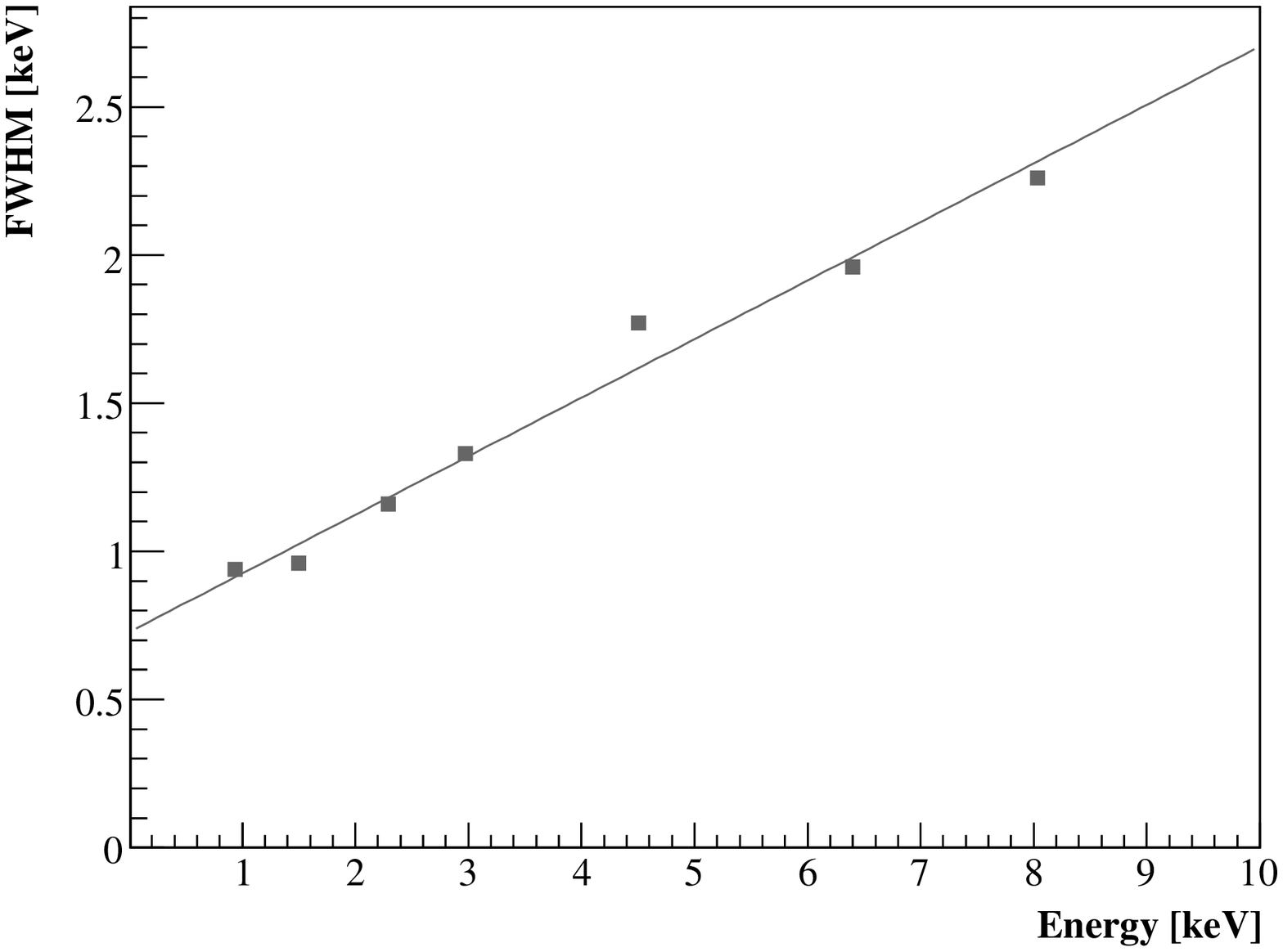}
    \end{minipage}
  \end{center}
  \caption{Left: Peak position in arbitrary ADC units versus incident
    photon energy. The black points and the grey points were taken from two
    different data sets, showing the variation of the detector gain due to
    different atmospheric conditions. Right: Measurements of the energy
    resolution of the TPC detector in terms of full width at half maximum
    (FWHM) of the photo-peak.}
  \label{linearity}
\end{figure}
The PANTER data show also the linearity of the TPC response. The position
of the main peak versus energy for each measured PANTER energy point is
plotted in \fref{linearity} (left), and this verifies the linearity of the
detector gain. The points of each set (two different atmospheric pressures)
closely follow a straight line, so the linearity of the chamber response
has been demonstrated down to the lowest tested X-ray energies.

The run with the lowest available PANTER energy, $0.3\,\text{keV}$, proved
that the TPC was sensitive to these energies, although with a very low
efficiency (for this run a special lower trigger threshold was set in the
acquisition electronics). The linearity of the detector response is also
preserved down to these low energies. Finally, the TPC energy resolution
can be also extracted from these data.  \Fref{linearity} (right) shows the
resolution in terms of Full Width at Half Maximum (FWHM) versus energy.

\section{Data Treatment}
\label{sec:data-treatment}
\begin{figure}[ht]
  \begin{center}
    \begin{minipage}{0.49\textwidth}
      \includegraphics[width=1.1\textwidth]{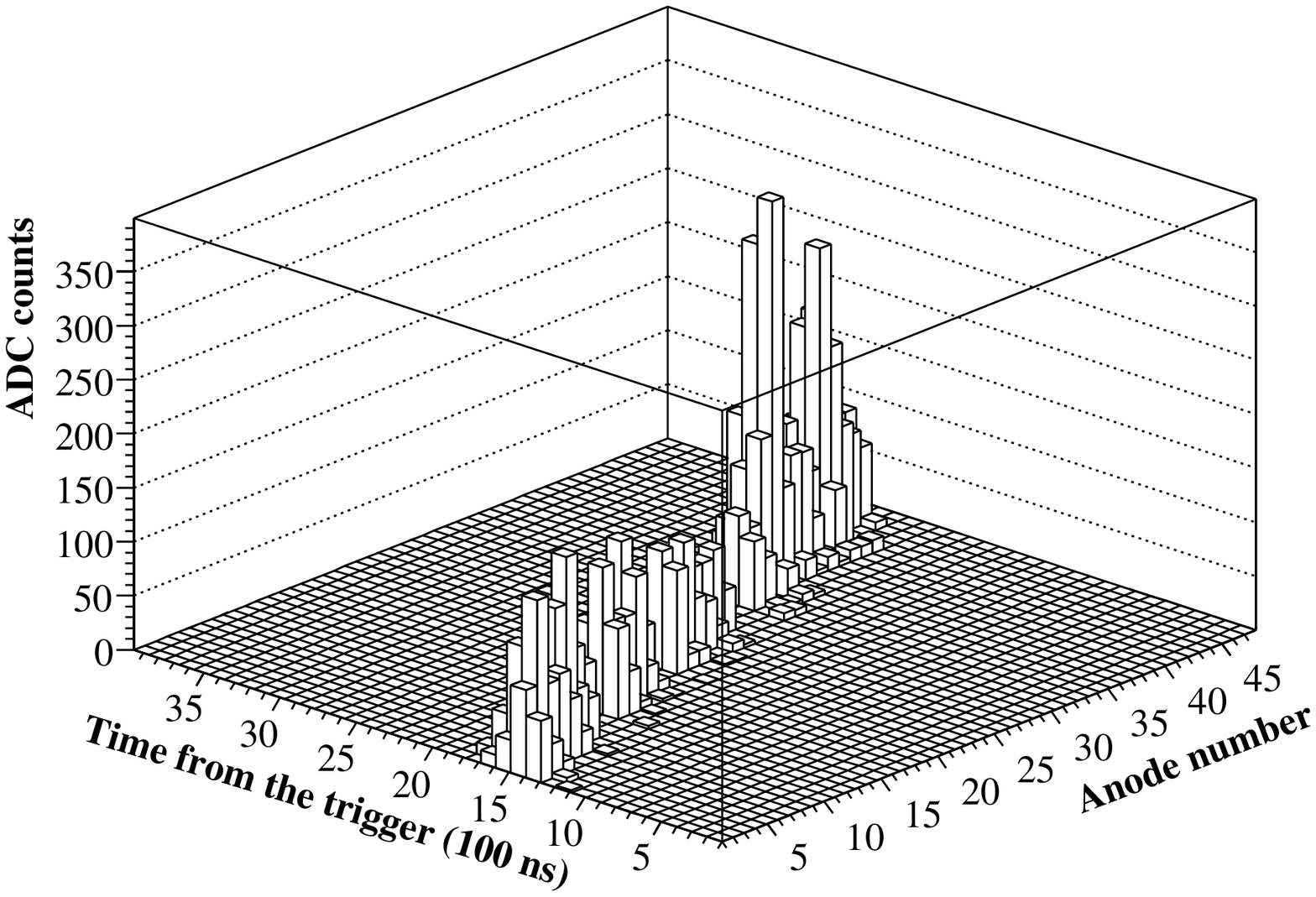}
    \end{minipage}
    \begin{minipage}{0.49\textwidth}
      \includegraphics[width=1.1\textwidth]{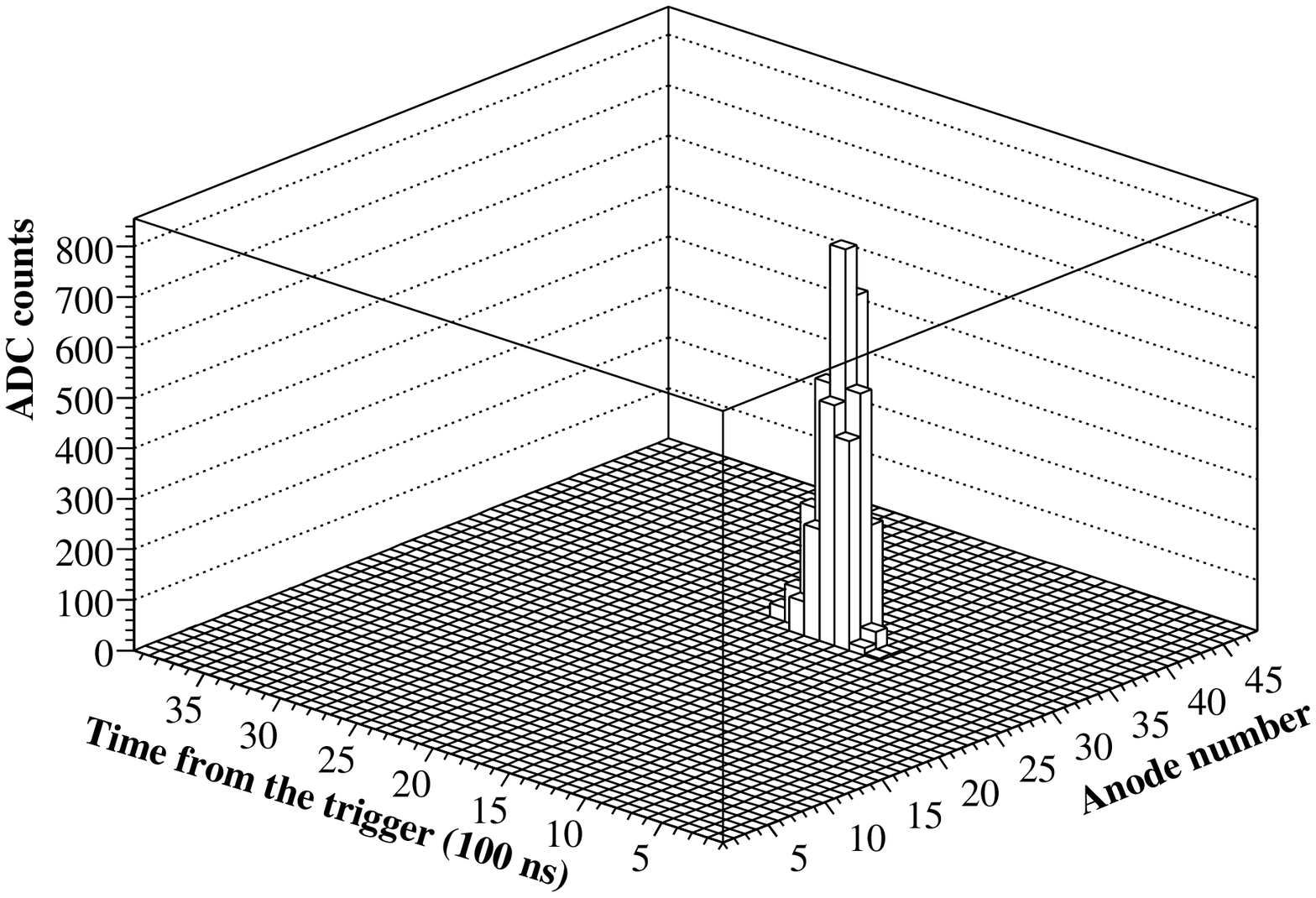}
    \end{minipage}
    \begin{minipage}{0.49\textwidth}
      \includegraphics[width=1.1\textwidth]{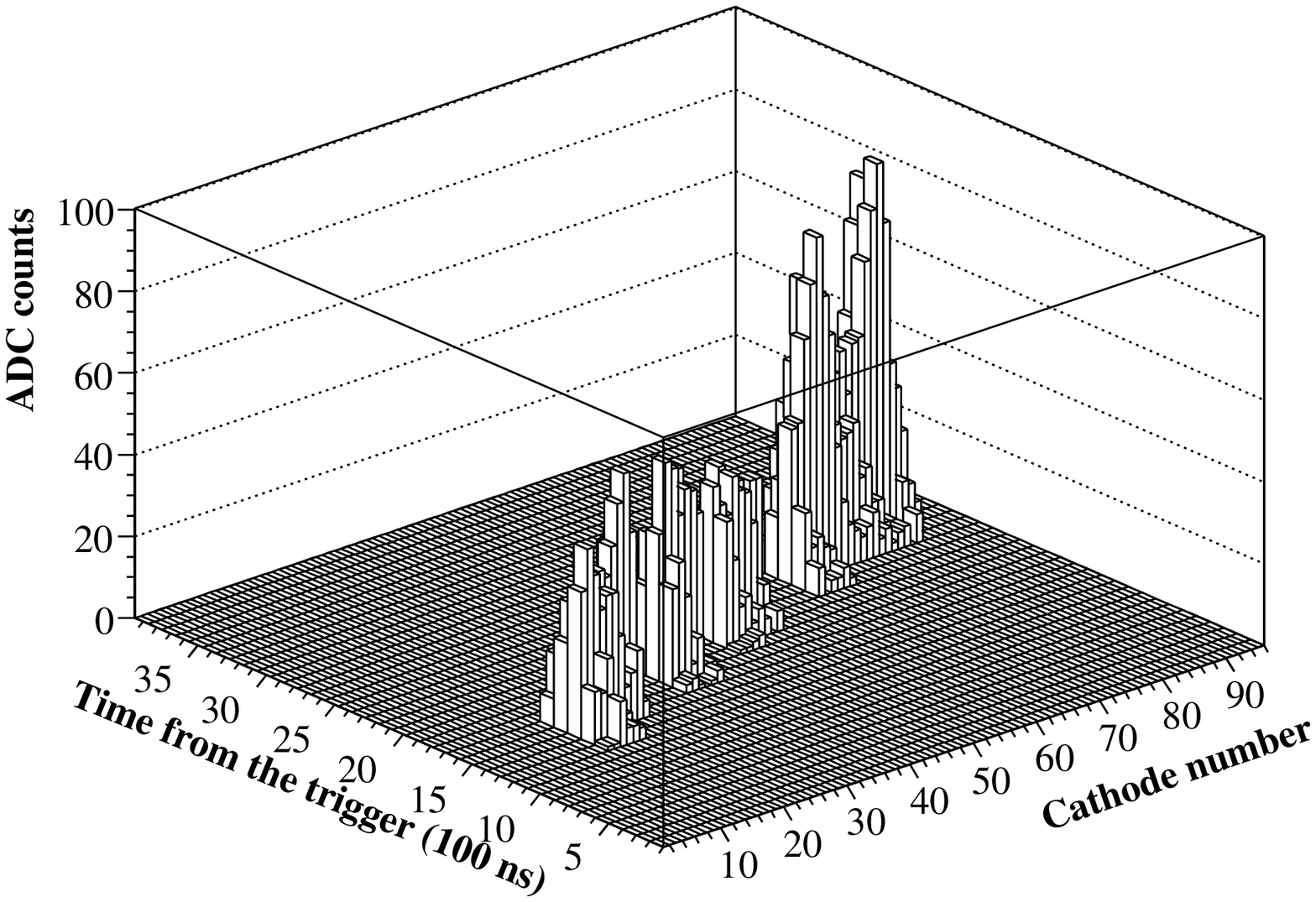}
    \end{minipage}
    \begin{minipage}{0.49\textwidth}
      \includegraphics[width=1.1\textwidth]{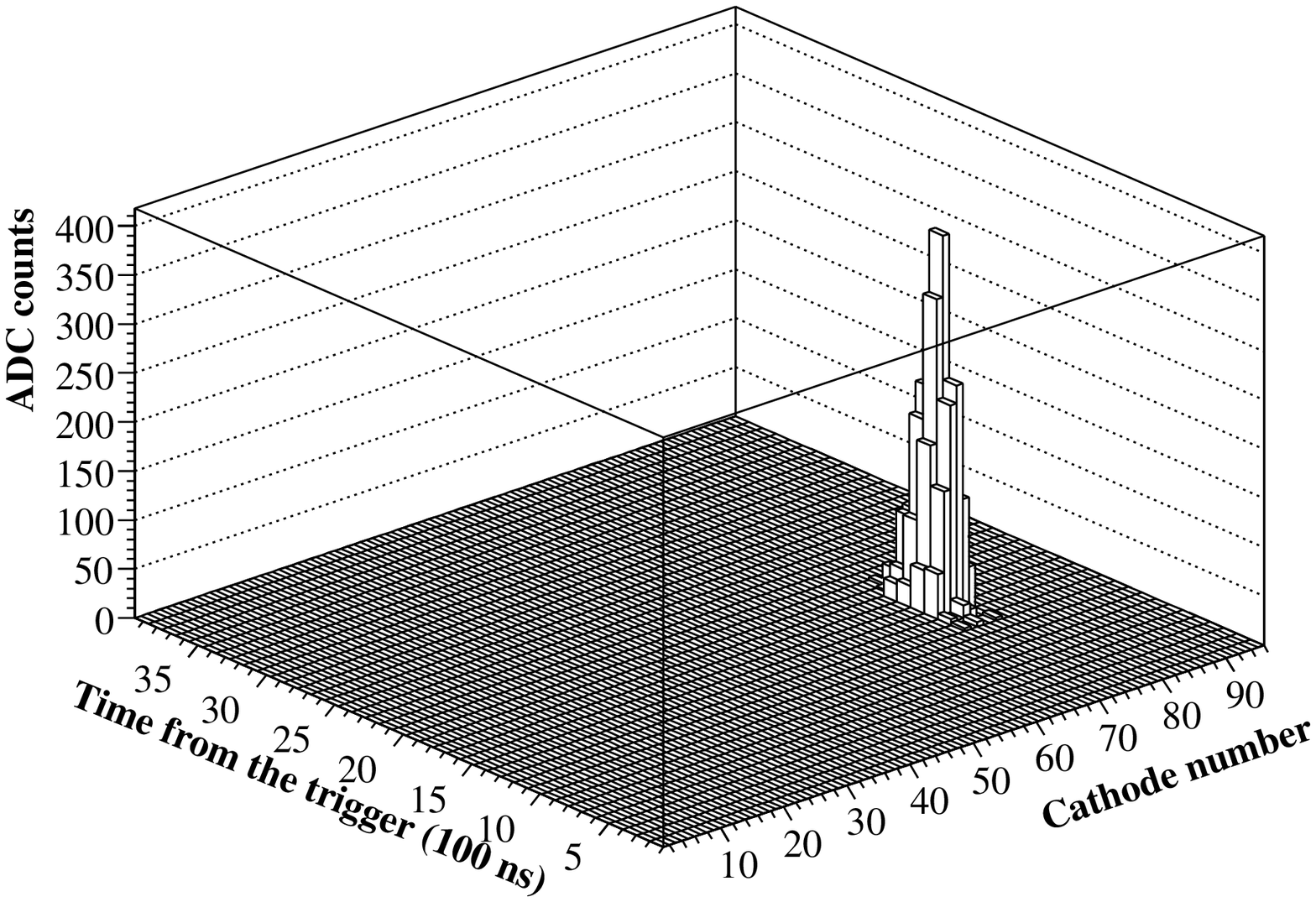}
    \end{minipage}
  \end{center}
  \caption{Time evolution of the charge pulse generated in each wire
    as recorded by the flash-ADCs. Top (bottom) plots correspond to what is
    collected in the anode (cathode) wires.  On the left a background event
    is plotted, while on the right a X-ray like event is shown.}
  \label{fig:anode}
\end{figure}
The first level data treatment of the CAST TPC is focused on the
identification of the almost point-like energy depositions produced by the
low energy X-rays in the conversion volume of the chamber.
\Fref{fig:anode} shows examples of such depositions for a background event
(left) and for an X-ray (right).  The plots on the top display the time
evolution of the charge pulse generated in each wire as recorded by the
flash-ADCs for the row of anodes, while on the bottom the same is shown for
the row of cathode wires. The very characteristic profile of an X-ray event
provides the framework for a selective analysis.

A first characterization of these raw data generated by the flash-ADCs is
performed depending on the spread of the signal. For analysis purposes, a
``cluster'' is defined as a set of charge pulses (hits) gathered on several
contiguous anode wires (anode cluster) as well as on several contiguous
cathode wires (cathode cluster) \footnote{For analysis purposes we
  distinguish clusters of anode hits and clusters of cathode hits
  separately. Obviously, events containing one single physical cluster must
  contain one anode cluster and one cathode cluster.}.  It has been proven
experimentally that X-rays of the energies of interest in our chamber
normally produce a single-cluster event firing between $1$ and $3$ anode
wires and between $2$ and $8$ cathode wires with a time difference between
contiguous hits less than $50\,\text{ns}$ (see \fref{fig:anode}). The
spread of the anode signal is mainly due to the diffusion of the electron
cloud along the drift distance and to a lesser extent to the initial energy
deposition (higher energies give larger initial ionization clouds). The
larger spread in the cathode wires is due to the development of the
avalanche process along the sense (anode) wires. In addition, to be
considered as a cluster, the total charge must exceed a minimum charge
threshold. This is done to avoid the improbable effect of correlated noise
in contiguous wires producing spurious clusters that would affect the
efficiency.  Therefore, the entire information for one raw data event can
now be reduced to a small set of numerical parameters, including:
\begin{itemize}
\item {Number of clusters} in the event and whether they are in the
  anode or cathode wire plane. This number provides information on the
  signal spread.
\item {Multiplicity} of every cluster, i.e., number of hits comprising the
  cluster.
\item {Total cluster charge} calculated by adding up the charge of
  every hit comprising the cluster.
\item {Cluster position} calculated by the charge-weighted mean of the
  position (wire number) of every hit comprising the cluster.
\item {Cluster time} (related to the trigger) calculated by the
  charge-weighted mean of the times of every hit composing the cluster.
\end{itemize}
Next, a set of software cuts are applied to reject events that are clearly
not produced by X-ray interactions. The first and most important one is the
requirement that there is one single anode cluster and one single cathode
cluster. It is then straightforward to match both of them to get the
2-dimensional position of the point-like event. A detailed list of further
conditions required on the cluster properties is given in table \ref{cuts}.
\begin{table}[t] 
  \centering \footnotesize
  \caption {\label{cuts}\it{List of software cuts applied to the CAST TPC
      data.}} 
  \begin{indented}
  \item[]\begin{tabular}{cc} \hline \hline
      \bf{cut} & \bf{condition} \\ \hline \\
      anode multiplicity & number of anode hits \\
      & between 1 and 3 (both inclusive) \\ \\
      cathode multiplicity & number of cathode hits \\
      & between 2 and 8 (both inclusive) \\ \\
      anode--cathode time difference & time between anode and cathode cluster\\
      & in the range $-0.15$ to $0.02$ $\mu\text{s}$\\\\
      no saturation & no hit reaching the upper part \\ & of the flash-ADC dynamical range\\\\
      anode--cathode charge ratio & around $1.85$, but slightly energy dependent\\\\
      fiducial cut & only events whose 2-D coordinates are inside \\ & the
      windows facing the magnet bores \\ \\ \hline
    \end{tabular}
  \end{indented}
\end{table}
This set of cuts is a minimal choice designed to reject a large portion of
background events while minimally reducing the efficiency of the detector.
Typically its application reduces the background by approximately two
orders of magnitude with respect to the raw trigger rate.  The loss of
efficiency generated by the software cuts has been carefully measured using
the PANTER data as described above, in particular regarding the loss of
X-rays events with split energy deposition (2 or more cluster events) due
to the argon escape peak.

\begin{figure}
  \begin{center}
    \includegraphics[width=0.8\textwidth]{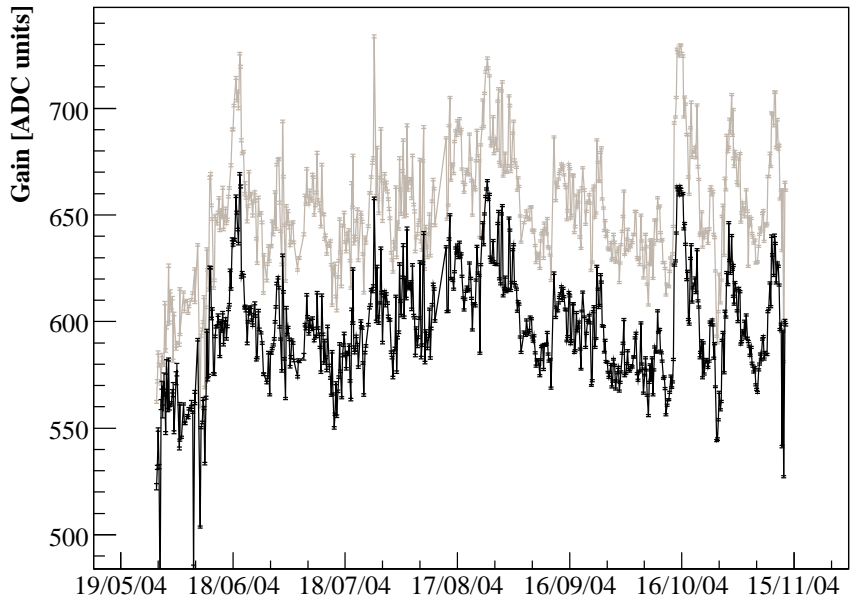}
  \end{center}
  \caption{Gain evolution during 2004 data taking in the two windows of
    the TPC, given by the ADC channel corresponding to the position of the
    Mn-K$\alpha$ line of an $^{55}${\rm Fe} calibration source.}
  \label{fig:gain}
\end{figure}

The cluster energy is obtained by adding up the calibrated flash-ADC output
for every constituent hit.  In 2003 data taking, the gain variations of the
gas in the chamber were characterized by continuous measurements of its
pressure and temperature. In 2004 the introduction of a stepping motor
allowed fully automated calibration runs with a $^{55}$Fe source every 6
hours as described in section~\ref{sec:data-acqu-hardw}. Therefore,
measurements of the gas gain for each window were performed and used to
calibrate the flash-ADCs.  \Fref{fig:gain} shows the evolution of the
measured gain during the 2004 data taking period.
A difference of the gain values for the two windows was observed and this
was taken into account in the data analysis. This difference results from
geometrical imperfections in the wire arrangement that produce a gain drift
along the anodes.  Slight wire-to-wire variations due to mechanical
imperfections or adjustments of the preamplifiers were also observed and
further corrected by calibrating the gain of every wire independently. The
energy spectrum of the events that have passed all cuts is then computed
for two distinct classes: the first for all X-ray events collected during
the axion-sensitive periods (when the magnet is pointing towards the Sun)
and the second for the background spectrum.  The X-rays coming from
conversion of solar axions should appear as an excess in the first spectrum
when compared to the second one.

\section{Detector Performance}
\label{sec:detector-performance}
\subsection{2003 Data Taking}
Although the data taking period of 2003 lasted for about 6 months, much of
this time corresponds to commissioning operation, periods when data taking
was temporarily stopped due to specific technical interventions in the
experiment (related for example to the problems with the mechanical
structure) or in the TPC itself (replacement of leaky windows, for example)
or periods when data was taken but they did not pass the quality
requirements regarding homogeneity of operation, due to the relatively
frequent interruptions in the experiment, mainly due to magnet quenches,
and to a lesser extent to episodes of electronic noise pick-up in the
detector.

\begin{figure}
  \begin{center}
    \includegraphics[width=0.75\textwidth]{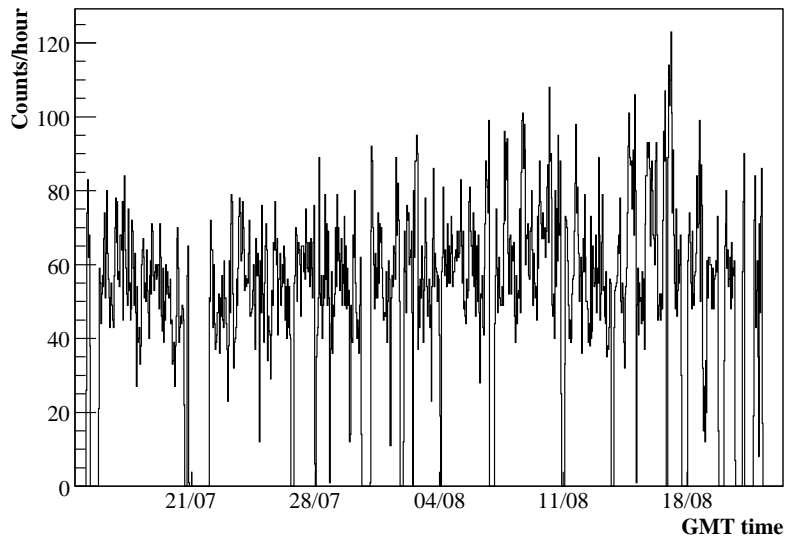} 
  \end{center}
  \caption{Measured count rate in the $3$ to $7\,{\rm keV}$ energy range in
    units of counts per hour as observed by the TPC detector during July
    and August 2003 after applying cuts.}
  \label{fig:rate-set6}
\end{figure}

As a result, the effective amount of data qualified for analysis obtained
by the TPC in 2003 was $\sim783\,\text{h}$, all of them concentrated in the
months of July and August. Out of these data, $\sim63\,\text{h}$ (9\% of
the total) were taken when the magnet was tracking the Sun. The stability
and continuity of operation of the detector during this time can be
appreciated in \fref{fig:rate-set6}, where the counting rate is
represented.
Regarding the performance of the TPC setup, the main problem preventing a
continuous operation beyond the period signaled was the deterioration of
the leak tightness of the X-ray windows beyond the magnet vacuum
requirements. This strongly motivated the development of the differential
pumping system previously described, and its installation at the beginning
of the 2004 data taking period.

On the other hand, during 2003 the TPC operated with a reduced shielding
consisting only of the innermost copper cage previously presented and the
N$_2$ flow. With this reduced shielding, the observed background suffered
from a strong dependence on the magnet position, caused by the relatively
large spatial movements at the far end of the magnet, which resulted, as
previously mentioned, in appreciably different environmental radioactivity
levels. This effect prevented a straightforward, background-subtracting,
analysis. To correct for this systematic effect an effective background
spectrum was constructed only from the background data taken in magnet
positions where Sun tracking had been performed and this was weighted
accordingly with the relative exposure of the tracking data.
\begin{figure}
  \begin{center}
    \includegraphics[width=0.75\textwidth]{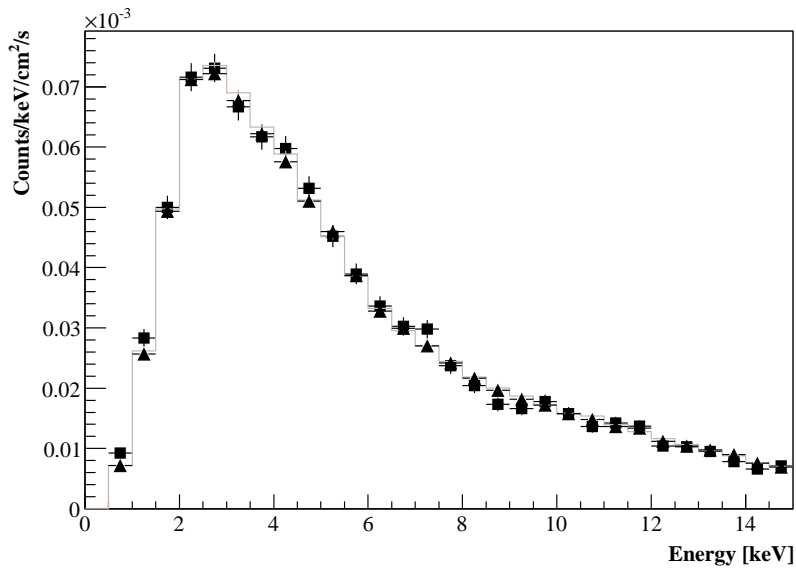} 
  \end{center}
  \caption{Three different spectra are shown: the total measured background
    spectrum (triangles), the weighted background spectrum (squares), and
    the tracking spectrum \citeaffixed{andriamonje:07a}{solid line, see
      also }.}
  \label{fig:fondo-pesado}
\end{figure}

In practice, we distinguish the differential spectrum of the complete
background data, ${\rm d}N_{\rm b}/{\rm d}E$ from the partial background
spectra $({\rm d}N_{\rm b}/{\rm d}E)_i$ corresponding to the data gathered
in the ``cell'' $i$ of the magnet position plane (azimuth-altitude). The
effective background spectrum to be subtracted from the tracking data that
will not suffer from this systematic effect regarding magnet position was
defined as:
\begin{equation}
  \label{eq:fondo-pesado}
  \left( \frac{{\rm d}N_{\rm b}}{{\rm d}E}\right)_{\rm eff}=\frac{\sum_i ({\rm
    d}N_{\rm b}/{\rm d}E)_i
    \epsilon_i}{\sum_i \epsilon_i}
\end{equation}
where $\epsilon_i$ is the exposure of the tracking data for magnet
positions in the cell $i$. The more cells used, the smaller the systematic
error will be when considering the effective background, but on the other
hand the higher the statistical errors due to the poor available statistics
in each of the cells.  For the 2003 data, it was seen that a grid of
$3\times3$ cells is sufficient to reduce this effect well below the
statistical error in the tracking data. In \fref{fig:fondo-pesado} both the
total background spectrum and the one obtained after the above weighting
are shown in comparison to the tracking spectrum.

\subsection{2004 Data Taking}
\begin{figure}
  \begin{center}
    \includegraphics[width=0.85\textwidth]{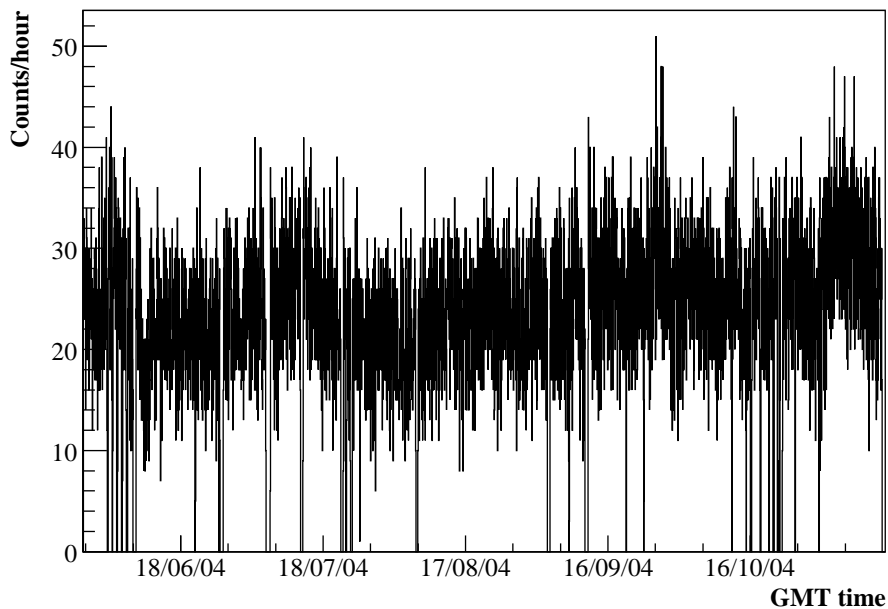} 
  \end{center}
  \caption{Measured count rate in the $3$ to $7\,{\rm keV}$ energy range  in
    units of counts per day as observed by the TPC detector during 2004
    data taking period.}
  \label{fig:rate-set9}
\end{figure}

In 2004, with the differential pumping system installed and the general
improvements in the CAST experiment, the whole TPC setup worked safely
without interruption for the full duration of the data taking period (six
months). The collected data are of very high quality, due to the
installation of the shielding and the detector stability.
\Fref{fig:rate-set9} shows the rate of filtered events between $3$ and
$7\,\text{keV}$ collected during this period.
Data were taken during $\sim 151$ effective days, of which 9~days ($\sim
6\%$) correspond to axion sensitive conditions.

\begin{figure}
  \begin{minipage}{0.49\textwidth}
    \begin{center}
      \includegraphics[width=1.12\textwidth]{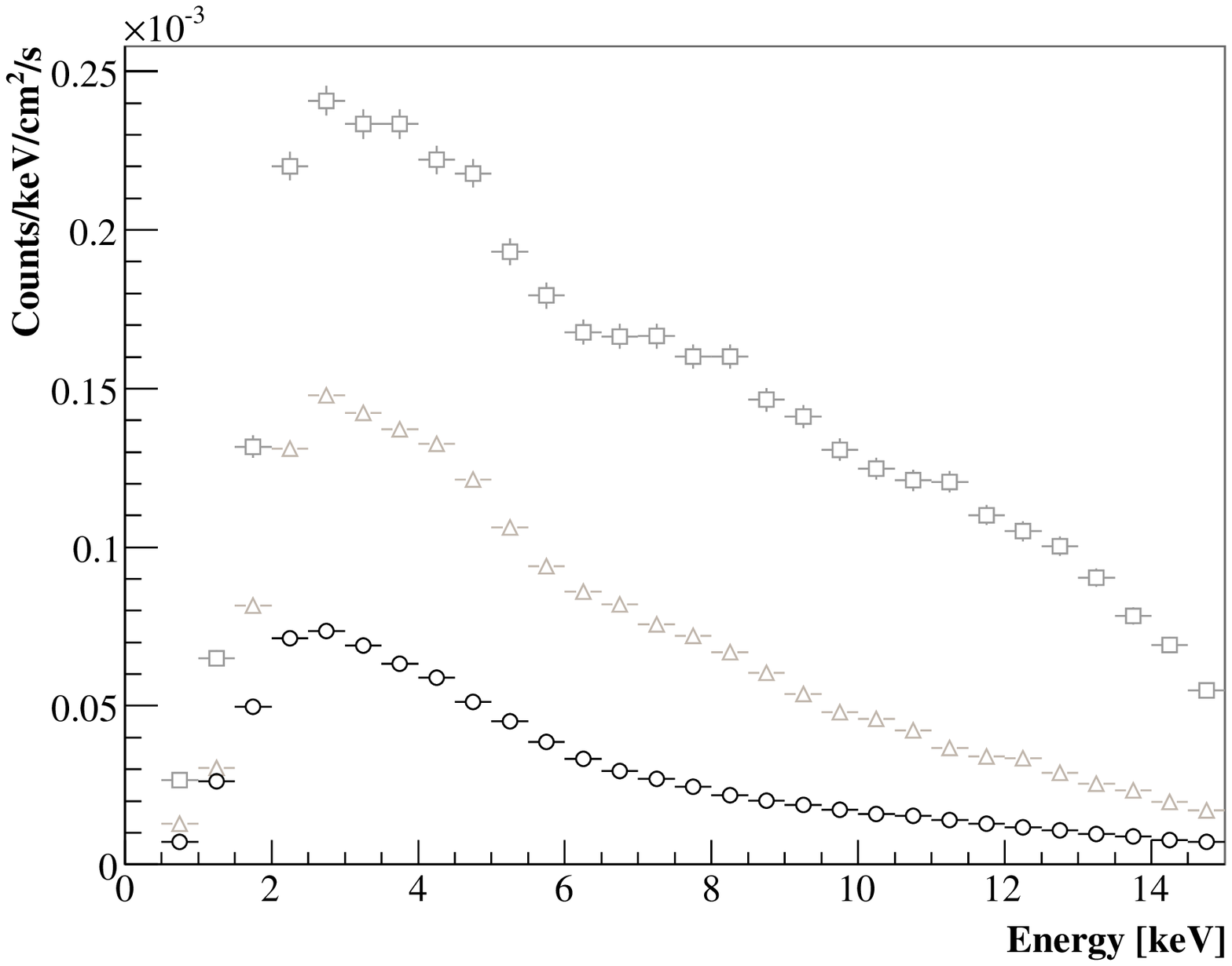}
    \end{center}
  \end{minipage}
  \hfill
  \begin{minipage}{0.49\textwidth}
    \begin{center}
      \includegraphics[width=1.05\textwidth]{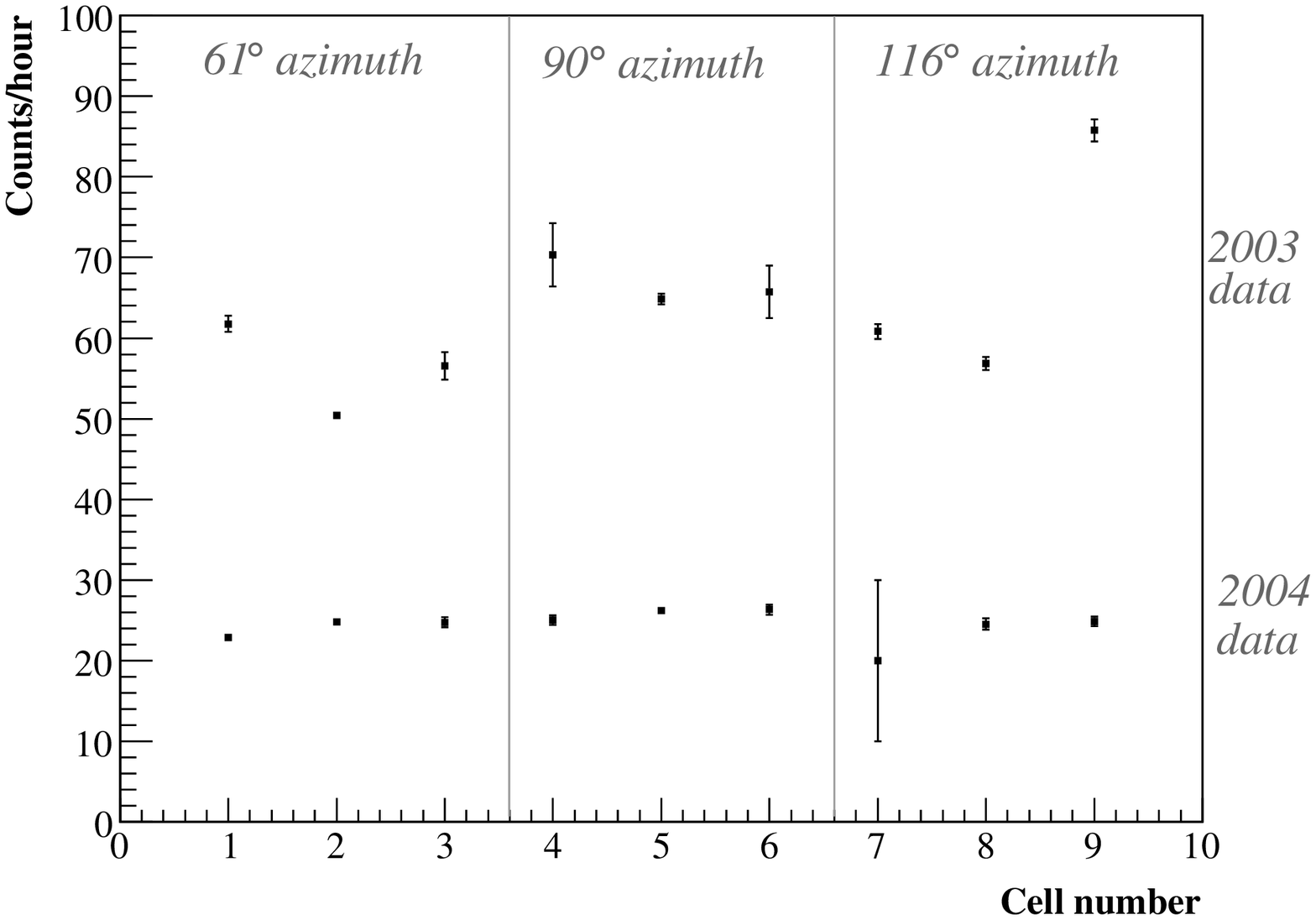}
    \end{center}
  \end{minipage}
  \caption{Left: Background spectra for the TPC detector measured under
    different shielding conditions. Right: Comparison between 2003 (top)
    and 2004 background data (bottom) measured for different magnet
    orientations, i.e., different detector location relative to the
    environment of the experiment area. The cell numbers indicate 3
    different horizontal positions and 3 different vertical positions.}
  \label{fig:positions}    \label{background2}
\end{figure}

With the shielding installed around the chamber, the TPC background level
between 1 and 10~keV was $4.10\times10^{-5}\,\ccmskev$, a factor of $\sim
4.3$ below the level reached by the TPC with no shielding. This reduction
increases with energy (reduction factor of $\sim 6.4$ in the $6$ to
$10\,\text{keV}$ range). The observed background energy spectra for both
setups with and without shielding are shown in the left panel of
\fref{background2}, as well as for an intermediate configuration with only
the copper box and N$_2$ flux.

Regarding the performance of the muon veto installed as part of the
shielding, it was observed, as expected, that the events in coincidence
with the muon veto are mostly rejected by the software cuts, since the muon
leaves a very distinctive signature in the TPC. The amount of events
rejected by anti-coincidence with the veto and not previously rejected by
the software cuts represent only 4\% of the final background.  This small
population is due to muon-induced events leaving a point-like energy
deposit in the TPC, like neutrons produced by the muons in the shielding.
 
In order to have a better control of the spatial background
inhomogeneities, background data were always taken in well defined
positions in the experimental hall.  For this purpose, an $3\times3$
$x$--$y$ grid was defined, where $x(y)$ is the TPC horizontal (vertical)
coordinate in the experimental hall. In this way, 9 cells cover the
different TPC positions, and the background level was measured in each of
them. The right panel of \Fref{fig:positions} shows the result of the
measurements for both the 2003 and 2004 data.  During 2003 the TPC was
shielded only with the copper box and the N$_2$ flux and differences in the
background rate due to the proximity of a concrete wall are clearly seen.
On the other hand, for the 2004 data the overall effect of the shielding
can be observed, both in the reduction and also in the clear homogenization
of the overall background level.

\section{Conclusion}
The CAST TPC detector, built to search for solar axions, has been presented
in detail. A design option of $10\,\text{cm}$ maximum drift length with a
conventional multi-wire readout plane has been implemented to provide the
experimental requirements of high sensitivity to X-rays from the magnet
bores. Some innovative features of the detector design include two thin
X-ray windows to allow photons coming from the magnet bores to enter the
detector with very small losses, a differential pumping system to couple
the chamber to the magnet while maintaining the stringent requirements of
leak rate towards the magnet vacuum, and a low background shielding built
around the detector to reduce the background and to make it more uniform in
space. A detailed characterization of the detector in the PANTER facility
in Munich has been performed, proving an adequate response linearity, a low
energy threshold and an overall (hardware plus software) efficiency of 62\%
for photons coming from conversion of solar axions. The imaging
capabilities of the multi-wire readout enable a high background
discrimination which, together with the radio-purity of the detector
materials and the detector shielding, results in a background rate of only
$4.10\times10^{-5}\,\ccmskev$ between $1$ and $10\,\text{keV}$. The whole
system has been proved to be robust and stable during the full data taking
of CAST phase~I, allowing for a reliable background subtraction and
therefore contributing to the overall CAST limit on axion-photon coupling
and mass \cite{zioutas:05a}.

\ack 

This work has been performed in the CAST collaboration. We thank our
colleagues for their support. We would like to express here our gratitude
to the group of the Laboratorio de F\'isica Nuclear y Altas Energ\'ias of
the Zaragoza University for material radio-purity measurements.
Furthermore, the authors acknowledge the helpful discussions within the
network on direct dark matter detection of the ILIAS integrating activity
(Contract number: RII3-CT-2003-506222).

\section*{References}
\bibliography{mnemonic,xmm,cast,conferences,detback,detector,general}
\bibliographystyle{jphysicsB}

\end{document}